\documentclass[10pt,twoside,twocolumn]{article}

\usepackage{amsmath}
\usepackage[latin1]{inputenc}
\usepackage[spanish,english]{babel}

\begin{document}

\title{The relation between momentum conservation and Newton's third law
revisited}
\vspace{-14mm}
\author{Rodolfo A. Diaz\thanks{%
radiazs@unal.edu.co}, William J. Herrera\thanks{%
jherreraw@unal.edu.co} \\
Universidad Nacional de Colombia, Departamento de F\'{\i}sica, Bogot\'{a}%
-Colombia.}
\date{}
\maketitle
\vspace{-11mm}
\begin{abstract}
\emph{Under certain conditions usually fulfilled in classical mechanics, the
principle of conservation of linear momentum and Newton's third law are
equivalent. However, the demonstration of this fact is usually incomplete in
textbooks. We shall show here that to demonstrate the equivalence, we
require the explicit use of the principle of superposition contained in Newton's second law. On the other hand, under some additional conditions the
combined laws of conservation of linear and angular momentum, are equivalent to Newton's third law with central forces. The conditions for such
equivalence apply in many scenarios of classical mechanics; once again the
principle of superposition contained in Newton's second law is the clue.}


\end{abstract}

\vspace{-6mm}

\selectlanguage{spanish}

\begin{abstract}
\emph{
Bajo ciertas condiciones que se cumplen con frecuencia en la mecánica clásica, los principios de conservación del momento lineal y la tercera ley de Newton son equivalentes. No obstante, la demostración de este hecho está usualmente incompleta en los textos sobre el tema. En este artículo se demuestra dicha equivalencia, para lo cual requerimos del uso explícito del principio de superposición contenido en la segunda ley de Newton. Por otro lado, bajo algunas condiciones adicionales las leyes combinadas de conservación del momento angular y del momento lineal, son equivalentes a la tercera ley de Newton con fuerzas centrales. Las condiciones para esta última equivalencia también son válidas en muchos escenarios de la mecánica clásica; una vez más el principio de superposición contenido en la segunda ley de Newton es la clave para la demostración.}

\noindent \textbf{Palabras clave}: \emph{Leyes de Newton, conservación del momento lineal, conservación del momento angular.}
\end{abstract}

\vspace{-1mm}

\selectlanguage{english}

A very important part of the foundations of classical mechanics lies on one hand in Newton's third law or, on the other hand, on the principle of conservation of linear momentum. Thus, the link between both approaches is of greatest interest in constructing the formalism of classical Physics. Commom texts of mechanics \cite{mecanica}, usually state that Newton's third
law automatically leads to the principle of conservation of linear momentum.
However, the converse is also true under certain conditions; the proof in
reverse order is either absent or restricted to systems of two particles in
most textbooks. \setcounter{footnote}{2} We start from the statement
of the principle of linear momentum conservation for a closed and isolated
system of particles\footnote{%
For a closed system, we mean a system in which particles are the same at all
times, i.e. no particle interchange occurs with the surroundings.}, with respect
to a certain inertial frame%
\begin{equation}
\mathbf{P}_{1}+\mathbf{P}_{2}+\ldots +\mathbf{P}_{n}=constant.  \label{mom}
\end{equation}%
Deriving with respect to time, gives%
\begin{equation}
\frac{d\mathbf{P}_{1}}{dt}+\frac{d\mathbf{P}_{2}}{dt}+\ldots +\frac{d\mathbf{%
P}_{n}}{dt}=0 ,  \label{derive}
\end{equation}%
but according to Newton's second law, $\frac{d\mathbf{P}_{i}}{dt}$
refers to the total force applied to the $i-$th particle. Further, since
there are no external forces Eq. (\ref{derive}) becomes%
\begin{eqnarray}
\sum_{j\neq 1}^{n}\mathbf{F}_{1j}+\sum_{j\neq 2}^{n}\mathbf{F}_{2j}+\ldots
+\sum_{j\neq n}^{n}\mathbf{F}_{nj} &=&0  \notag \\
\Rightarrow \hspace{4mm}\sum_{i=1}^{n}\sum_{j\neq i}^{n}\mathbf{F}_{ij}&=&0 .
\label{fund eq}
\end{eqnarray}%
Where $\mathbf{F}_{ij}$ stands for the force on the $i-$th particle due to
the $j-$th particle. In the case of two particles, Eq. (\ref{fund eq}) leads
to Newton's third law automatically. However, in the case of an arbitrary
number of particles, Newton's third law is only a sufficient condition
in this step. The proof of necessity is the one that is usually absent in
textbooks.

In order to prove the necessity, we shall use the principle of superposition
stated in Newton's second law. Considering a system of $n$ particles, let us take
a couple of particles $k$ and $l$. They undergo the force of each other $%
\mathbf{F}_{kl}$ and $\mathbf{F}_{lk}$ respectively, plus the internal
forces due to the other particles of the system. However, according to the
second law, the forces $\mathbf{F}_{kl}$ and $\mathbf{F}_{lk}$ are not
altered by the presence of the rest of the forces (i.e. the other forces do
not interfere with them). Therefore if we withdrew the other particles of
the system leaving the particles $k$ and $l$ in the same position, the
forces $\mathbf{F}_{kl}$ and $\mathbf{F}_{lk}$ would be the same as those
when all particles were interacting. Now, after the withdrawal of the other
particles, our system consists of two isolated particles for which the third
law is evident. Hence $\mathbf{F}_{kl}=-\mathbf{F}_{lk}$. We proceed in
the same way for all the pairs of particles and obtain that $\mathbf{F}%
_{ij}=-\mathbf{F}_{ji}$ for all $i,j$ in the system. Observe that the proof
of necessity requires the use of the principle of superposition contained in
Newton's second law. Since we have demonstrated the necessity and
sufficiency, we have proved the equivalence. Notwithstanding, this
equivalence is based on many implicit assumptions

\begin{enumerate}
\item \emph{Newton's second law is valid}: As is well known, in
scenarios such as quantum mechanics the concept of force is not meaningful
any more.

\item \emph{The time runs in the same way for all inertial observers}: We
have used this condition since in the time derivative of Eq. (\ref{derive})
we do not mention what inertial system we have used to measure the time.
Besides, this condition is necessary to assume that the force is equal in
all inertial systems.

\item \emph{All the momentum of the system is carried by the particles}: In
this approach we are ignoring the possible storage or transmission of
momentum from the fields generated by the interactions (see discussion in
Ref. \cite{ajp49}).

\item \emph{The signals transmitting the interactions travel instantaneously}%
: In Eq. (\ref{mom}), each momentum $\mathbf{P_{i}}$ is supposed to be
measured at the same time. If any particle of the system changes its
momentum at the time $t$; then to preserve the law of conservation of
momentum (at the time $t$), it is necessary for the rest of the particles to
change their momenta \textbf{at the same time}, in such a way that they
cancel the change of momentum caused by the $i-th\ $particle. This fact is
in turn related with the condition that all the momentum be carried by the
particles (mechanical momentum). In other words, the other particles must
learn of the change in momentum of the $i-$th particle \textbf{instantaneously.}
\end{enumerate}

\vspace{2mm}

As has been emphasized in the literature, even in the case in which all
these assumptions fail, the principle of momentum conservation is still held
while Newton's third law is not valid any more, from which follows
the advantage of formulating the empirical principles of classical mechanics
in terms of the concept of momentum. Even when the assumptions given above
are satisfied, the formulation in terms of momentum is advantageous \cite%
{mom}. Nevertheless, we emphasize that under the conditions cited above, Newton's third law is equivalent to the principle of conservation of linear momentum, but the complete proof of that statement requires the
principle of superposition of forces established by Newton's second law.

On the other hand, by a similar argument we can show the equivalence of
combined conservation of linear and angular momentum with Newton's third
law with central forces. Starting from the conservation of angular momentum
for a closed isolated system with respect to an inertial frame

\begin{equation}
\mathbf{L}=\mathbf{L}_{1}\mathbf{+L}_{2}\mathbf{+...+L}_{n}=constant
\label{ang}
\end{equation}%
and deriving this equation we find 
\begin{equation*}
\frac{d\mathbf{L}_{1}}{dt}\mathbf{+}\frac{d\mathbf{L}_{2}}{dt}\mathbf{+...+}%
\frac{d\mathbf{L}_{n}}{dt}=0 .
\end{equation*}%
From the definition of $\mathbf{L}_{i}$ and taking into account that the
system is isolated, the derivative of the total angular momentum reads%
\begin{equation*}
\frac{d\mathbf{L}}{dt}=\sum_{i=1}^{n}\left( \mathbf{r}_{i}\times \sum_{j\neq
i}^{n}\mathbf{F}_{ij}\right) =0 .
\end{equation*}%
Under the assumption $\mathbf{F}_{ij}=-\mathbf{F}_{ji}$ (obtained from the
conservation of linear momentum) we can show the following
identity by induction%
\begin{equation*}
\sum_{i=1}^{n}\left( \mathbf{r}_{i}\times \sum_{j\neq i}^{n}\mathbf{F}%
_{ij}\right) =\sum_{i=1}^{n-1}\sum_{j>i}^{n}\left[ \left( \mathbf{r}_{i}-%
\mathbf{r}_{j}\right) \times \mathbf{F}_{ij}\right] .
\end{equation*}%
Clearly,  Newton's third law with central forces (i.e. $\mathbf{F}_{ij}=-%
\mathbf{F}_{ji}$ and $\left( \mathbf{r}_{i}-\mathbf{r}_{j}\right) \ $%
parallel to$\ \mathbf{F}_{ij}$) is a sufficient condition for the conservation of angular
and linear momentum (we shall refer to the third law with
central forces as the \emph{strong version of Newton's third law}
henceforth). To prove necessity we resort again to the argument of
isolating one pair of particles $k,l\ $without changing their positions.
Since this two-particle system is now isolated, its total angular momentum
must be constant, and remembering that the forces $\mathbf{F}_{kl}=-\mathbf{F%
}_{lk}$ have not changed either, we have%
\begin{equation*}
\frac{d\mathbf{L}\left( two\ particles\right) }{dt}=\left( \mathbf{r}_{k}-%
\mathbf{r}_{l}\right) \times \mathbf{F}_{kl}=0 .
\end{equation*}%
Now, since both particles have different positions and we are assuming that
they are interacting ($\mathbf{F}_{kl}\neq 0$)\footnote{%
The case of $\mathbf{F}_{kl}=0$ could be considered as a trivial realization
of the strong version of Newton's third law.}, we obtain that $\left( 
\mathbf{r}_{k}-\mathbf{r}_{l}\right) $ must be parallel to $\mathbf{F}_{kl}$%
. We can proceed in a similar way for all possible pairs of particles in the
system. In this case we have used the combined laws of conservation of linear and angular
momentum because Newton's third law in its weak version was
assumed since the beginning. Of course, the conditions for this equivalence
to hold are those cited above, but with analogous assumptions for angular momentum as well.

As before, when the conditions for this equivalence fail, the laws of
conservation of linear and angular momentum are still valid, while the
strong version of Newton's third law no longer holds.

An important issue arises when we consider non-isolated systems, since we
have assumed that the system is isolated throughout this treatment. If we add
external forces, once again the principle of superposition states that the
internal forces do not interfere with them, and so Newton's third law is
maintained. A similar argument holds for possible external torques and the
Newton's third law in its strong version.

In conclusion, we have proved that under certain conditions the principle of
linear momentum conservation is equivalent to the weak version of the
Newton's third law. Analogously, under similar conditions, the combined laws
of conservation of linear and angular momentum are equivalent to the strong
version of Newton's third law. We emphasize that for both demonstrations
we should invoke the principle of superposition contained in Newton's
second law\footnote{%
Of course, Newton's first law is also in the background of all this
treatment, by assuming the existence of inertial frames in which the other
laws of Newton and the conservation of momenta are valid.}. Finally, it is worth
mentioning that the suppositions neccesary to obtain these equivalences are implicit in the
original formalism of classical mechanics \cite{laws}. Therefore, such
equivalences deserve more attention, at least until reaching
relativity, quantum mechanics or classical (quantum) field theories.

The authors wish to thank Dr. H\'{e}ctor M\'{u}nera for useful discussions and
Dr. Victor Tapia for his suggestions concerning the bibliography.

\end{document}